\documentclass[pra,superscriptaddress,twocolumn]{revtex4-2}

\usepackage{enumerate}
\usepackage{amsfonts,amssymb,amsmath}
\usepackage[]{graphics,graphicx,epsfig}
\usepackage{amsthm}
\usepackage{graphicx}
\usepackage{dcolumn}
\usepackage{natbib}
\usepackage{color}
\usepackage{multirow}
\usepackage{ulem}
\usepackage{bm}
\usepackage{url}

\usepackage[colorlinks = true, citecolor=blue, linkcolor=blue, urlcolor=blue]{hyperref}
\newcommand*{\fullref}[1]{\hyperref[{#1}]{\autoref*{#1} \nameref*{#1}}}

\begin{document}

\title{Quasi-probabilistic Bit Erasure Causes Bell Non-Locality} 

\author{Kelvin Onggadinata}
\affiliation{Centre for Quantum Technologies,
National University of Singapore, 3 Science Drive 2, 117543 Singapore,
Singapore}
\affiliation{Department of Physics,
National University of Singapore, 3 Science Drive 2, 117543 Singapore,
Singapore}

\author{Pawe{\l} Kurzy{\'n}ski}
\affiliation{ Institute of Spintronics and Quantum Information, Faculty of Physics, Adam Mickiewicz University, Uniwersytetu Pozna{\'n}skiego 2, 61-614 Pozna\'n, Poland}
\affiliation{Centre for Quantum Technologies,
National University of Singapore, 3 Science Drive 2, 117543 Singapore,
Singapore}

\author{Dagomir Kaszlikowski}
\email{phykd@nus.edu.sg}
\affiliation{Centre for Quantum Technologies,
National University of Singapore, 3 Science Drive 2, 117543 Singapore,
Singapore}
\affiliation{Department of Physics,
National University of Singapore, 3 Science Drive 2, 117543 Singapore,
Singapore}

\date{\today}

\begin{abstract}
We show that a maximal violation of the Bell-CHSH inequality for two entangled qubits, i.e., Bell non-locality, is a direct consequence of a local bit erasure by means of a quasi-stochastic process, i.e., a stochastic process in which some transition probabilities are negative.
\end{abstract}

\maketitle

\section{Introduction}
The orthodox way of looking at quantum systems is Hilbert space mathematics. Wigner showed an alternative that later was turned into a full-fledged theory of {\it frames} \cite{ferrie2008frame, ferrie2009framed}. Frames use quasi-probabilities in the intermediate stages of calculations, always ending up with positive probabilities describing measurement outcomes in the lab, freeing us from a cumbersome task of negative probability interpretations. It is entirely up to you what description you choose as they are fully equivalent as many worlds are to Copenhagen interpretation of quantum theory. In this paper we show that if you choose the latter to describe a typical Bell-type experiment \cite{bell1964einstein,clauser1969propesed,RevModPhys.86.419} with two entangled qubits, you come across a strange phenomena where a peculiar choice of lossy, local information processing denies local reality of measurement outcomes, i.e., Bell non-locality emerges. 

Since frames are an alternative description of atomic world, we can pretend we live in a universe where David Hilbert was never born and Andrei Kolmogorov entertained negative probabilities that he subsequently used to explain quantum mechanics of an arbitrary number of qubits. However, to make the reader's life easier we will relate frames to Hilbert space when necessary.

\section{A qubit}

Here is how one can represent one and two qubits as quasi-stochastic systems within frames formalism. In this paper we chose one particular frame instead of dealing with a general frame formalism. This is similar to choosing, for instance, a convenient reference frame in classical mechanics to solve a problem at hand. We discuss different possible choices in the Discussion section of the paper.     

We start with a single qubit represented by a density matrix in Hilbert space
\begin{equation}
    \rho=\frac{1}{2}(\openone + s_x \sigma_x + s_y \sigma_y +s_z \sigma_z)\, ,
\end{equation}
where $s_i$ are the components of the Bloch vector $\vec{s}$ ($|\vec{s}| \leq 1$) and $\sigma_i$ are Pauli operators. $\vec{s}$ provides a full description of the qubit's state.

In the quasi-probabilistic representation in the SIC-POVM frame, a single qubit is a system of two classical bits $a,a'=\pm 1$, always distributed with the following {\it positive} probability distribution
\begin{equation}
p(aa')=\frac{1}{4}(1+\vec{s}\cdot\hat{n}_{aa'})\, ,
\end{equation}
where $\vec{s}$ is the same Bloch vector as in the standard description and $\hat{n}_{aa'}=\frac{1}{\sqrt{3}}(a,a',aa')$. This distribution belongs to a proper subset, a 3D ball, of a classical probabilistic simplex with the vertices given by four pure probability distributions 
\begin{equation}
w_{bb'}(aa')=\frac{1}{4}(1+3\hat{n}_{bb'}\cdot\hat{n}_{aa'})=\delta_{ab}\delta_{a'b'}\, .
\end{equation}
Bloch vector $\vec{s}$, in this approach, is simply a convenient way to encapsulate information about average values of bits $a,a'$: $\langle A\rangle = \frac{s_x}{\sqrt{3}}$, $\langle A'\rangle = \frac{s_y}{\sqrt{3}}$ and $\langle AA'\rangle=\frac{s_z}{\sqrt{3}}$, i.e.,

\begin{equation}
    p(aa')=\frac{1}{4}(1+\frac{s_x}{\sqrt{3}}a+\frac{s_y}{\sqrt{3}}a'+\frac{s_z}{\sqrt{3}}aa')\, .
\end{equation}

Dynamics of a qubit is now any quasi-stochastic process $S(bb'|aa')$ ($\sum_{bb'}S(bb'|aa')=1$ but some entries $S(bb'|aa')$ can be negative) that keeps $\vec{s}$ inside the 3D ball. There is a specific class of such processes $S(bb'|aa')=\frac{1}{4}(1+3\hat{n}_{bb'}\cdot O\vec{n}_{aa'})$ that simply rotate $\vec{s}: \vec{s}\rightarrow O\vec{s}$, where $O$ is an arbitrary orthogonal matrix -- let us call such processes {\it rotations} for brevity. They correspond to unitaries in Hilbert space. Rotations form a continuous group of quasi-stochastic processes that are also quasi-bistochastic, i.e., $\sum_{aa'}S(bb'|aa')=1$. They can be perfectly reversed, i.e., there always exists $S^{-1}(cc'|bb')$ such that $\sum_{bb'}S^{-1}(cc'|bb')S(bb'|aa')=\delta_{ca}\delta_{c'a'}$. This is not possible for positive bistochastic processes unless they are permutations. All the other allowed processes are affine transformations of $\vec{s}: \vec{s}\rightarrow \Lambda \vec{s}+\vec{s}_0$ and they correspond to Krauss channels in Hilbert space \cite{nielsen2010quantum}.

A measurement on a qubit is another quasi-stochastic process. Here we have a multitude of options. Let us start with the simplest one: $R(bb'|aa')=\frac{1}{4}(1+3\hat{n}_{bb'}\cdot\hat{n}_{aa'})=\delta_{ab}\delta_{a'b'}.$ This is a direct readout of two bits describing the qubit, or differently, a stochastic process whose transition probabilities are one for $aa'\rightarrow aa'$ and zero otherwise. Another interesting example is a process $\eta(b|aa')=\frac{1}{2}(1+3 b \hat{a}\cdot\hat{n}_{aa'})$ where $\hat{a}$ is a real 3D unit vector ($|\hat{a}|=1$). Note that both processes yield positive probability distributions although the second one is a quasi-stochastic, irreversible process. We have 
\begin{equation}
p(b)=\sum_{aa'}\eta(b|aa')p(aa')=\frac{1}{2}(1+b\hat{a}\cdot\vec{s})\geq 0\, ,
\end{equation}
where $b$ is the measurement bit we observe in the laboratory. The process $\eta$ recovers a quantum measurement of the Pauli operator $\hat{a}\cdot\vec{\sigma}$, where $b$ is the measurement outcome. Moreover, it can be seen as a positive stochastic process controlled by a negative bit (nebit) as shown in \cite{onggadinata2023simulations}. Direct readout $R$ corresponds to a SIC-POVM measurement. 

We have many more options. Any quasi-stochastic process mapping two bit probability distribution $p(aa')$ to a positive probability distribution over bit strings $b_1b_2\dots b_N$ is permissible
\begin{equation}
M(b_1b_2\dots b_N|aa')=\frac{1}{2^N}(1+\vec{m}_{b_1b_2\dots b_N}\cdot\vec{w}_{aa'})\, ,
\end{equation}
where $\sum_{b_1b_2\dots b_N}\vec{m}_{b_1b_2\dots b_N}=0$ (probability normalisation condition) and $\vec{w}_{aa'}$ are some three dimensional vectors. We get
\begin{eqnarray}
&&p(b_1b_2\dots b_N)=\sum_{aa'}M(b_1b_2\dots b_N|aa')p(aa')\\
&&=\frac{1}{2^N}(1+\frac{1}{4} \vec{w}\cdot\vec{m}_{b_1b_2\dots b_N}+\vec{m}_{b_1b_2\dots b_N}\cdot(M\vec{s}))\, ,
\end{eqnarray}
where $\vec{w}=\sum_{aa'}\vec{w}_{aa'}$ and $M=\sum_{aa'}\vec{w}_{aa'}\vec{n}_{aa'}$ ($\vec{w}_{aa'}\vec{n}_{aa'}$ is a dyadic product). However, in this paper, only the processes $R$ and $\eta$ are of interest to us.

The first process is a direct readout of two bits describing a qubit. We get all the information that is available out there and no negative probabilities are involved. In some sense, this is the most {\it natural} measurement because it simply tells us what is ``out there". Statistics of this direct readout give us vector $\vec{s}$ that fully determines qubit's state, i.e., the bit averages $\langle A\rangle, \langle A'\rangle$ and $\langle AA'\rangle$ or, if one wishes, the Bloch vector $\vec{s}$.

The second process is more interesting as it corresponds to the orthodox and most controversial von Neumann projective measurement. In the frames formalism this seems to be reflected in the necessity to use negative probabilities to implement it. Moreover, it erases one bit of information about the qubit's state, $aa'\rightarrow b$, in the sense that having only statistics of measurement bit, $b$, cannot determine the qubit's state, i.e., $\vec{s}$ is only partially known. In comparison, the direct readout $R$ does not erase any information, it is a passive observation, bearing resemblance to measurements in classical physics.  

\section{Two qubits}

In the orthodox language the state of two qubits, Alice's and Bob's, is represented by
\begin{eqnarray}
& &\rho_{AB}=  \\
& &\frac{1}{4}\left(\openone\otimes\openone + \vec{s}_A \cdot \vec{\sigma}\otimes\openone + \openone\otimes\vec{s}_B \cdot\vec{\sigma} \sum_{i,j=x,y,z}T_{ij} \sigma_i\otimes\sigma_j\right) \nonumber
\end{eqnarray}
where $\vec{s}_A$ and $\vec{s}_B$ are local Bloch vectors of the corresponding qubits and $T_{ij}$ are elements of the correlation matrix $T$.

In the quasi-probabilistic description this becomes a {\it positive} probability distribution 
\begin{equation}
p(aa',bb')=\frac{1}{16}(1+\vec{s}_A\cdot\hat{n}_{aa'}+\vec{s}_B\cdot\hat{n}_{bb'}+\hat{n}_{aa'}\cdot T\hat{n}_{bb'})\, .
\label{twoq}
\end{equation}
 Such distributions are proper subsets of the classical simplex spanned by pure four-bit distributions, similarly to a single qubit distributions. It is obvious that $\vec{s}_A$ describes qubit $A$ and $\vec{s}_B$ qubit B because $\sum_{bb'}p(aa',bb')$ should refer to qubit $A$ and $\sum_{aa'}p(aa',bb')$ to qubit $B$. $T$ contains information about correlations between qubits $A$ and $B$, i.e., between bit pairs $aa'$ and $bb'$. 

Dynamics of two qubits is any quasi-stochastic process $S(aa',bb'|\alpha \alpha',\beta \beta')$ that takes $p(\alpha \alpha', \beta \beta')$ to $p(aa',bb')$, preserving the structure of (\ref{twoq}). Note that local, single qubit dynamics are valid choices. 

A particular state $p_0$ is of interest to us:
\begin{equation}
p_{0}(aa',bb')=\frac{1}{16}(1-\hat{n}_{aa'}\cdot\hat{n}_{bb'})\, .
\end{equation}
This state, corresponding to the singlet state, has an interesting symmetry: if Alice and Bob choose the same local rotations (quasi-stochastic processes defined above) it does not change because $\hat{n}_{aa'}\cdot\hat{n}_{bb'}$ is invariant under same rotations. Another property of this state is that each local distribution for Alice and Bob is a pure white noise yet it has a large amount of correlations -- matching pairs of bits on both sides never appear, i.e., $p(aa',aa')=0$.  

In this paper we are only interested in the situation where bits $aa'$ belong to Alice spatially separated from Bob who owns bits $bb'$. Thus we already know the mathematics of Alice and Bob's measurements as discussed in the preceding sections.

\section{Bell-CHSH violations}

What follows next is a direct translation of the Bell-CHSH scenario \cite{clauser1969propesed} to frames formalism. Alice and Bob apply two randomly chosen rotations $S^A_1, S^A_2$ and $S^B_1, S^B_2$, respectively, to the commonly shared state $p_0$. They obtain four different positive probability distributions $p_0^{ij}(aa';bb')$, $i,j=1,2$:
\begin{eqnarray}
&&p_0^{ij}(aa',bb') \nonumber \\
&& \qquad = \sum_{\alpha\alpha',\beta\beta'} S^A_i(aa'|\alpha\alpha')S^B_j(bb'|\beta\beta')p_0(\alpha\alpha',\beta\beta')\\
&& \qquad =\frac{1}{16}(1-O^A_i\hat{n}_{aa'}\cdot O^B_j\hat{n}_{bb'}).
\end{eqnarray}
This is the orthodox Bell-CHSH type scenario used to test local reality in quantum theory \cite{bell1964einstein,clauser1969propesed}. Local reality was first introduced by EPR \cite{einstein1935quantum} and it can be interpreted that measurement outcomes exist before they are measured. Observation is a mere book keeping of what is already out there and that Alice's measurements do not instantenously affect spatially separated Bob. Technically, this amounts to an existence of a joint, positive probability distribution of all measurement outcomes \cite{fine1982hidden} obtained in the experiment.  

\subsection{Direct readout $R$}

Suppose Alice and Bob perform the direct readout $R$ of their bits for each state $p_0^{ij}$ obtaining two sets of bit pairs $a_ia'_i$ and $b_jb'_j$ ($i,j=1,2$) each. Are these bit pairs elements of local reality? As mentioned before, this is mathematically equivalent to finding a joint, positive probability distribution $p(a_1a_1',a_2a_2';b_1b_1',b_2b_2')$ returning $p_0^{ij}$ as marginals. 

The answer is `yes' and the proof follows Fine's method \cite{fine1982hidden}, where we explicitly construct a joint positive probability distribution $p(a_1a_1',a_2a_2';b_1b_1',b_2b_2')$:
\begin{eqnarray}
&&p(a_1a_1',a_2a_2';b_1b_1',b_2b_2')=\nonumber\\
&&\quad \frac{p(a_1a_1';b_1b_1',b_2b_2')p(a_2a_2';b_1b_1',b_2b_2')}{p(b_1b_1',b_2b_2')}\,,
\end{eqnarray}
where 
\begin{eqnarray}
&&p(a_1a_1';b_1b_1',b_2b_2') = \nonumber\\
&&\qquad \frac{1}{64}(1-\hat{n}_{a_1a_1'}\cdot\hat{n}_{b_1b_1'})(1-\hat{n}_{a_1a_1'}\cdot\hat{n}_{b_2b_2'})\, ,\\
&&p(a_2a_2';b_1b_1',b_2b_2') = \nonumber\\
&&\qquad \frac{1}{64}(1-\hat{n}_{a_2a_2'}\cdot\hat{n}_{b_1b_1'})(1-\hat{n}_{a_2a_2'}\cdot\hat{n}_{b_2b_2'})\, ,
\end{eqnarray}
and 
\begin{equation}
p(b_1b_1',b_2b_2')= \frac{1}{16}(1-\frac{1}{3}\hat{n}_{b_1b_1'}\cdot\hat{n}_{b_2b_2'})\, .
\end{equation}
The validity of this explicit construction relies on the fact that
\begin{eqnarray}
\sum_{a_1a_1'}p(a_1a_1';b_1b_1',b_2b_2') &=&  \sum_{a_2a_2'}p(a_2a_2';b_1b_1',b_2b_2')\nonumber\\
&=& p(b_1b_1',b_2b_2')
\end{eqnarray}
and that all intermediate probability distributions are positive. Therefore, Alice's bits $aa'$ and Bob's bits $bb'$, popping up in this experiment, exist objectively.  

\subsection{Measurement $\eta$ }

What happens if Alice and Bob measure $\eta$? Since they first perform their respective rotations they can chose measurements along the same direction, say $\hat{z}$, getting the following probability distributions
\begin{eqnarray}
   q_{ij}(ab)&=&\sum_{\alpha\alpha',\beta\beta'} \eta(a|\alpha\alpha')\eta(b|\beta\beta')p_0^{ij}(\alpha\alpha',\beta\beta') \nonumber \\ &=&\frac{1}{2}(1+abO^A_i\hat{z}\cdot O^B_j\hat{z}).
\end{eqnarray}
It is known that for a certain choice of rotations $O_A^i,O_B^j$ there is no positive joint probability distribution returning $q_{ij}(ab)$ as marginals and thus pairs of measurement bits $ab$ do not exist objectively in this experiment, they are `created' during the measurement\cite{bell1964einstein, clauser1969propesed,RevModPhys.86.419}. Moreover, the measurement $\eta$, for optimal rotations, maximally violates the CHSH inequality up to $2\sqrt{2}$. Objective and real bits can only reach $2$. 

\section{Discussion}

One can choose to represent physical qubits and their dynamics as quasi-stochastic systems: positive probability distributions of bits as qubits' states and quasi-stochastic processes as qubits' dynamics and measurements. This picture is in one-to-one correspondence with the Hilbert space orthodoxy via frames \cite{ferrie2008frame, ferrie2009framed} and thus it is a matter of preference as, for instance, is many worlds interpretation vs. Copenhagen interpretation of quantum mechanics. 

We have shown that in the chosen quasi-stochastic picture, a non-existence of local reality in an experiment with two qubits and two sets of local random measurements on a state $p_0$ can be directly attributed to a classical bit erasure with the help of quasi-probabilities (or, equivalently, with the help of a nebit) via a measurement process $\eta$. Without this erasure, when one uses a direct readout $R$ instead, local reality for measurement outcomes does exist or, alternatively, bits $aa'$ are objectively real in this experiment. 

It was shown before that Bell non-locality can be a result of negative probabilities \cite{Abramsky_2011}. In particular, it was shown that  Bell non-locality can emerge due to quasi-stochastic transformations of positive probability distributions \cite{PhysRevLett.111.170403}. However, it was not identified which particular quasi-stochastic processes are responsible for this emergence. The main observation of this work is that a quasi-stochastic reversible dynamics of a qubit is not sufficient to deny local reality since such a dynamics followed by the direct readout $R$ admits it. Of course, it is possible that if we test local reality with readout $R$, where Alice and Bob have more random rotations instead of two, qubit's bits $aa'$ may lose their objective existence as well.  

These findings suggest that quantum supremacy, which occurs in Bell non-locality scenarios \cite{RevModPhys.86.419}, may happen due to information erasure during measurements at the end of a quantum algorithm execution. This is because in the quasi-probabilistic picture, a quantum algorithm is a quasi-bistochastic operation taking a multi-qubit positive probability distribution $p_{in}(a_1a_1',a_2a_2',\dots,a_Na_N')$ and transforming it to another positive distribution $p_{out}(a_1a_1',a_2a_2',\dots,a_Na_N')$. If the findings of this paper translate to $N$ qubits then, only bit erasure measurements $\eta$ carried on the individual qubits can introduce non-classical effects. This is only a speculation at this moment and a future research direction.    

\section*{Acknowledgements}

This research is supported by the National Research Foundation, Singapore, and A*STAR under the CQT Bridging Grant. PK is supported by the Polish National Science Centre (NCN) under the Maestro Grant no. DEC-2019/34/A/ST2/00081. 

\bibliography{ref}

\end{document}